\documentclass[a4paper,11pt]{article}
\usepackage{pos}

\usepackage{textcomp}
\usepackage{amsfonts} 
\usepackage{amsmath} 
\usepackage{slashed}
\usepackage{pifont}
\usepackage{multirow,lscape}
\usepackage{array}
\usepackage{subcaption}
\usepackage{verbatim}
\usepackage{bm}
\usepackage{comment}
\usepackage{xcolor}
\usepackage{url} 

\graphicspath{{./fig/}}

\providecommand{\wbar}[1]{\overline#1}

\providecommand{\mate}[3]{\langle#1\lvert#2\rvert#3\rangle}

\renewcommand{\Re}{\mathrm{Re}\,}
\renewcommand{\Im}{\mathrm{Im}\,}

\providecommand{\GeV}{\;\mathrm{GeV}}
\providecommand{\MeV}{\;\mathrm{MeV}}

\definecolor{HLBlue}{HTML}{6599FF}
\definecolor{HLOrange}{HTML}{FF6600}

\newcommand{\BK}{\hat{B}_{K}}
\newcommand{\Vcb}{|V_{cb}|}

\newcommand{\Vus}{|V_{us}|}
\newcommand{\Vud}{|V_{ud}|}

\newcommand{\eps}{\varepsilon}

\newcommand{\epsK}{\varepsilon_{K}}

\newcommand{\BtoDstp}{\bar{B} \to D^{(\ast)} \ell \bar{\nu}}
\newcommand{\BtoDst}{\bar{B} \to D^\ast \ell \bar{\nu}}

\newcommand{\red}[1]{\textcolor{red}{#1}} 
\newcommand{\blue}[1]{\textcolor{blue}{#1}}

\begin{document}
\title{2024 Update on $\epsK$ with lattice QCD inputs}
\ShortTitle{$\epsK$ with lattice QCD inputs}

\author*[a]{Seungyeob Jwa}
\author[a]{Jeehun Kim}
\author[a]{Sunghee Kim}
\author[b]{Sunkyu Lee}
\author[a,1]{Weonjong Lee}
\author[c]{Jaehoon Leem}
\author[d]{Sungwoo Park}


\affiliation[a]{Lattice Gauge Theory Research Center, CTP, and FPRD,
  Department of Physics and Astronomy, \\
  Seoul National University,
  Seoul 08826, South Korea} 

\affiliation[b]{Center for Precision Neutrino Research (CPNR), 
  Department of Physics, \\ Chonnam National
  University, Yongbong-ro 77, Puk-gu, Gwangju 61186, Korea}

\affiliation[c]{Computational Science and Engineering Team,
  Innovation Center, Samsung Electronics, Hwaseong,
  Gyeonggi-do 18448, South Korea.}

\affiliation[d]{Lawrence Livermore National Lab, 7000 East Ave,
      Livermore, CA 94550, USA}

\emailAdd{wlee@snu.ac.kr}

\note{For the SWME collaboration}

\abstract{
  We report recent progress on $\epsK$ evaluated directly from the
  standard model (SM) with lattice QCD inputs such as $\BK$, exclusive
  $\Vcb$, $\Vus$, $\Vud$, $\xi_0$, $\xi_2$, $\xi_\text{LD}$, $f_K$,
  and $m_c$.
  We find that the standard model with exclusive $\Vcb$ and lattice
  QCD inputs describes only $2/3 \cong 65\%$ of the experimental value
  of $|\epsK|$ and does not explain its remaining 35\%, which
  represents a strong tension in $|\epsK|$ at the $5.1\sigma \sim
  4.1\sigma$ level between the SM theory and experiment.
  We also find that this tension disappears when we use the inclusive
  value of $\Vcb$ obtained using the heavy quark expansion based on
  the QCD sum rule approach.
  We also report results for $|\epsK|$ obtained using the
  Brod-Gorbahn-Stamou (BGS) method for $\eta_i$ of $u-t$ unitarity,
  which leads to even a stronger tension of $5.7\sigma \sim 4.2\sigma$
  with lattice QCD inputs.  }
\FullConference{The 41st International Symposium on Lattice Field
  Theory (LATTICE2024)\\ 28 July - 3 August 2024\\ Liverpool, UK\\}

\maketitle

%
%
\section{Introduction}
This paper is an update of our previous reports \cite{ Jwa:2023uuz,
  Lee:2023lxz, Lee:2021crz, Kim:2019vic, Bailey:2018feb,
  Bailey:2015tba, Bailey:2018aks, Jang:2017ieg, Bailey:2015frw}.
We report recent progress in the determination of $|\epsK|$ with
updated inputs from lattice QCD.
Updated input parameters include $\lambda$, $\bar{\rho}$,
$\bar{\eta}$, exclusive $\Vcb$, $\Vus$, $\Vud$, $\Vus/\Vud$, $M_W$,
$m_c(m_c)$, and $M_t$ (the pole mass of top quarks).

Here we adopt the same color convention as that in our previous papers
\cite{ Jwa:2023uuz, Lee:2023lxz, Lee:2021crz, Kim:2019vic,
  Bailey:2018feb, Bailey:2015tba, Bailey:2018aks, Jang:2017ieg,
  Bailey:2015frw} in Tables
\ref{tab:input-Vus-Vud}--\ref{tab:input-eta}.
We use \red{\texttt{red}} for new input data used to evaluate $\epsK$.
We use \blue{\texttt{blue}} for new input data which is not used.
%

%
%

%
%
%
\section{Input parameters: Wolfenstein parameters}
\label{sec:wp}
We summarize results for $\Vud$, $\Vus$, and $\dfrac{\Vus}{\Vud}$ from
lattice QCD in table \ref{tab:input-Vus-Vud}.
\begin{table}[h!]
  \begin{subtable}{0.57\linewidth}
    \renewcommand{\arraystretch}{1.1}
    \centering
    \resizebox{1.0\linewidth}{!}{
      \begin{tabular}{ l @{\quad} l @{\quad} l @{\quad} c }
        \hline\hline
        type  &  $\Vus$  & $\Vud$  & Ref.
        \\ \hline
        $N_f = 2+1+1$ & 0.22483(61)       & 0.97439(14)
        & FLAG-24 \cite{FlavourLatticeAveragingGroupFLAG:2024oxs}p79t20
        \\
        $N_f = 2+1$   & \blue{0.22481(58)} & \blue{0.97440(13)}
        & FLAG-24 \cite{FlavourLatticeAveragingGroupFLAG:2024oxs}p79t20
        \\ \hline\hline
      \end{tabular}
    } 
    \caption{$\Vus$ and $\Vud$ from lattice QCD}
    \label{tab:Vus+Vud}
  \end{subtable} 
  \hfill
  \begin{subtable}{0.42\linewidth}
    \renewcommand{\arraystretch}{1.1}
    \centering
    \resizebox{1.0\linewidth}{!}{
      \begin{tabular}{ l @{\qquad} l @{\qquad} c }
        \hline\hline
        type  &  $\Vus/\Vud$  & Ref.
        \\ \hline
        $f_{K^{\pm}}/f_{\pi^{\pm}}$ & \red{0.23126(50)}
        & FLAG-24 \cite{FlavourLatticeAveragingGroupFLAG:2024oxs}p75
        \\
        $f_K/f_\pi$                 & \red{0.23131(45)}
        & FLAG-24 \cite{FlavourLatticeAveragingGroupFLAG:2024oxs}p76
        \\ \hline\hline
      \end{tabular}
    } 
    \caption{$\Vus/\Vud$ from lattice QCD}
    \label{tab:Vus/Vud}
  \end{subtable} 
  \caption{ (\subref{tab:Vus+Vud}) $\Vus$ and $\Vud$
    (\subref{tab:Vus/Vud}) $\Vus/\Vud$. 
	}
  \label{tab:input-Vus-Vud}
\end{table}

\begin{align}
  \lambda &= \frac{ \Vus }{ \sqrt{ \Vud^2 + \Vus^2 } }
  = \frac{ r }{ \sqrt{ 1 + r^2 } }
  \,, \qquad\qquad
  r = \frac{\Vus}{\Vud}
  \label{eq:lambda-1}
\end{align}
Using Eq.~\eqref{eq:lambda-1}, we determine $\lambda$ from the ratio
$r = \Vus/\Vud$, since it has less error than that obtained directly
from $\Vus$ and $\Vud$.
We present results for $\lambda$ in Table \ref{tab:input-WP-eta}.
%
%
We also summarize recent update on $\bar\rho$ and $\bar\eta$ of
Wolfenstein parameters (WP) in Table \ref{tab:input-WP-eta}.
When we evaluate $|\epsK|$, we use the angle-only-fit (AOF) results in
Table \ref{tab:input-WP-eta} to avoid the unwanted correlations
between $(\epsK, \Vcb)$, and $(\bar\rho, \bar\eta)$, as explained in
Ref.~\cite{ Bailey:2018feb, Bailey:2015frw}.
We determine the parameter $A$ directly from $\Vcb$.
We also present results of the CKM-fitter \cite{ValeSilva:2024jml}
(2021) and the UTfit \cite{ UTfit:2022hsi, UTfit:2006vpt}(2022-2023) in
Table \ref{tab:input-WP-eta} for comparison.
\begin{table}[h!]
  \renewcommand{\arraystretch}{1.2}
  \resizebox{1.0\linewidth}{!}{
    \begin{tabular}{ @{\qquad} c @{\qquad} | @{\qquad} l @{\qquad} 
        l @{\qquad} | @{\qquad} l @{\qquad}  l @{\qquad} | @{\qquad}
        l @{\qquad} l @{\qquad} }
      \hline\hline
      WP
      & \multicolumn{2}{c}{CKMfitter}
      & \multicolumn{2}{c}{ UTfit}
      & \multicolumn{2}{c}{AOF}
      \\ \hline
      $\lambda$
      & $0.22498^{+0.00023}_{-0.00021}$         & \cite{ValeSilva:2024jml}
      & \blue{ 0.22519(83) }            & \cite{ UTfit:2022hsi, UTfit:2006vpt}
      & \red{  0.22536(42) }            & \cite{FlavourLatticeAveragingGroupFLAG:2024oxs}
      \\ \hline
      $\bar{\rho}$
      & $0.1562^{+0.0112}_{-0.0040}$        & \cite{ValeSilva:2024jml}
      & \blue{ 0.160(9) }            & \cite{ Bona:2024bue, UTfit:2006vpt}
      & \red{ 0.159(16) }            & \cite{ Bona:2024bue}
      \\ \hline
      $\bar{\eta}$
      & $0.3551^{+0.0051}_{-0.0057}$       & \cite{ValeSilva:2024jml}
      & \blue{ 0.346(9) }            & \cite{ Bona:2024bue, UTfit:2006vpt}
      & \red{ 0.339(10) }            & \cite{ Bona:2024bue}
      \\ \hline\hline
    \end{tabular}
  } 
  \caption{ Wolfenstein parameters }
  \label{tab:input-WP-eta}
\end{table}

%

%
%

%
%
\section{Input parameters: $\Vcb$}
\label{sec:Vcb}
We present recent results for exclusive $\Vcb$ and inclusive $\Vcb$ in
Table \ref{tab:Vcb}.
In Table \ref{tab:Vcb} (\subref{tab:ex-Vcb}), the results for
exclusive $\Vcb$ obtained by various groups of lattice QCD are
summarized: FNAL/MILC, FLAG, HFLAV, and HPQCD.
Here \texttt{BGL} denotes a kind of the parametrization methods for
data analysis \cite{Bailey:2018feb,Boyd:1997kz}, and \texttt{comb} represents combined
results from various groups for multiple decay channels.
They are consistent with one another within $1.0\sigma$ uncertainty.
We also present recent results for inclusive $\Vcb$ in Table
\ref{tab:Vcb} (\subref{tab:in-Vcb}).
There are a number of attempts to determine inclusive $\Vcb$ from
lattice QCD, but all of them at present belong to the category of
exploratory study rather than that of precision calculation \cite{
  Barone:2022gkn, Barone:2023tbl}. 
%
%
%
\begin{table}[t!]
  \begin{subtable}{1.0\linewidth}
    \renewcommand{\arraystretch}{1.2}
    \center
    \vspace*{-5mm}
    \resizebox{1.0\textwidth}{!}{
      \begin{tabular}{@{\qquad} l @{\qquad\qquad} l @{\qquad\qquad} l @{\qquad \qquad} l @{\qquad\qquad} l @{\qquad}}
        \hline\hline
        channel & value & method & ref & source \\ \hline
        $B\to D^* \ell \bar{\nu}$
        & \red{38.40(78)} & BGL
        & \cite{ FermilabLattice:2021cdg} p27e76 & FNAL/MILC-22
        \\
        ex-comb
        & \red{39.46(53)} & comb
        & \cite{FlavourLatticeAveragingGroupFLAG:2024oxs}p181e282  & FLAG-24
        \\
        ex-comb & \red{39.10(50)} & comb
        & \cite{ HeavyFlavorAveragingGroup:2022wzx} p120e221 & HFLAV-23  
        \\
        ex-comb & \red{39.03(56)(67)} & comb
        & \cite{Harrison:2023dzh} p24e50	& HPQCD-23  
        \\ \hline\hline
      \end{tabular}
    } 
    \caption{Exclusive $\Vcb$ in units of $1.0 \times 10^{-3}$. 
    }
    \label{tab:ex-Vcb}
  \end{subtable}  
  \begin{subtable}{1.0\linewidth}
    \renewcommand{\arraystretch}{1.2}
    \center
    \vspace*{+2mm}
    \resizebox{1.0\textwidth}{!}{
      \begin{tabular}{ @{\qquad} l @{\qquad\qquad\qquad} l @{\qquad\qquad\qquad\qquad} l @{\qquad\qquad\qquad} l @{\qquad} }
        \hline\hline
        scheme        & value         & ref  & source \\ \hline
        kinetic scheme & \blue{42.16(51)}
        & \cite{ Bordone:2021oof} p1 & Gambino-21
        \\
        1S scheme      & \red{41.98(45)}
        & \cite{ HeavyFlavorAveragingGroup:2022wzx} p108e200 & 1S-23 
        \\ \hline\hline
      \end{tabular}
    } 
    \caption{Inclusive $\Vcb$ in units of $1.0 \times 10^{-3}$.}
    \label{tab:in-Vcb}
  \end{subtable}
  \caption{ Results for (\subref{tab:ex-Vcb}) exclusive $\Vcb$ and
    (\subref{tab:in-Vcb}) inclusive $\Vcb$. The abbreviation p27e76 means
    Eq.~(76) on page 27.}
  \label{tab:Vcb}
    \vspace*{-3mm}
\end{table}
%
%
%

%

%
%

%
%
\section{Input parameter $\xi_0$}
The absorptive part of long distance effects on $\epsK$ is parametrized
by $\xi_0$.
\begin{align}
  \xi_0  &= \frac{\Im A_0}{\Re A_0}, \qquad
  \xi_2 = \frac{\Im A_2}{\Re A_2}, \qquad
  \Re \left(\frac{\eps'}{\eps} \right) =
  \frac{\omega}{\sqrt{2} |\eps_K|} (\xi_2 - \xi_0) \,.
  \label{eq:e'/e:xi0}
\end{align}
In lattice QCD, we can determine $\xi_0$ by two independent methods:
the direct and indirect methods.
In the direct method, one determines $\xi_0$ by combining the lattice
QCD results for $\Im A_0$ with experimental results for $\Re A_0$.
In the indirect method, one determines $\xi_0$ using
Eq.~\eqref{eq:e'/e:xi0} with lattice QCD results for $\xi_2$ combined
with experimental results for $\eps'/\eps$, $\epsK$, and $\omega$.
We summarize experimental results for $\Re A_0$ and $\Re A_2$,
lattice results for $\Im A_0$ and $\Im A_2$ calculated 
by RBC-UKQCD in Table~\ref{tab:xi0-sum}.
We also present results for $\xi_0$ in Table~\ref{tab:xi0-sum}.
Here we use the results of the indirect method for $\xi_0$ to evaluate
$\epsK$, since the total errors are much smaller than those of the
direct method.

\begin{table}[htbp]
  \begin{subtable}{1.0\linewidth}
    \renewcommand{\arraystretch}{1.2}
    \center
    \vspace*{-7mm}
    \resizebox{1.0\textwidth}{!}{
      \begin{tabular}{ @{\qquad} l @{\qquad} l @{\qquad\qquad} l @{\qquad\qquad} l @{\qquad} l @{\qquad} }
        \hline\hline
        parameter & method & value & Ref. & source \\ \hline
        $\Re A_0$ & exp & $3.3201(18) \times 10^{-7} \GeV$ &
        \cite{ Blum:2015ywa, RBC:2015gro,}  & NA
        \\
        $\Re A_2$ & exp & $1.4787(31) \times 10^{-8} \GeV$ &
        \cite{ Blum:2015ywa} & NA
        \\ \hline
        $\omega$ & exp & $0.04454(12)$ &
        \cite{ Blum:2015ywa} & NA
        \\ \hline
        $|\epsK|$ & exp & $2.228(11) \times 10^{-3}$ &
        \cite{ ParticleDataGroup:2024cfk} p285e13.46a & PDG-2024
        \\
        $\Re(\eps'/\eps)$ & exp & $1.66(23) \times 10^{-3}$ &
        \cite{ ParticleDataGroup:2024cfk} p285e13.46b & PDG-2024
        \\ \hline\hline
    %
  %
  %
        parameter & method & value ($\GeV$) & Ref. & source \\ \hline
        $\Im A_0$ & lattice/G-parity BC & \textcolor{red}{$-6.98(62)(144) \times 10^{-11}$} &
        \cite{ RBC:2020kdj} p4t1   & RBC-UK-2020 
        \\
        $\Im A_0$ & lattice/periodic BC & \textcolor{blue}{$-8.7(12)(26) \times 10^{-11}$} &
        \cite{ RBC:2023ynh} p30e70   & RBC-UK-2023 
        \\
        $\Im A_2$ & lattice/G-parity BC & \textcolor{red}{$-8.34(103) \times 10^{-13}$}  &
        \cite{ RBC:2020kdj} p31e90 & RBC-UK-2020
        \\
        $\Im A_2$ & lattice/periodic BC & \textcolor{blue}{$-5.91(13)(175) \times 10^{-13}$}  &
        \cite{ RBC:2023ynh} p30e68 & RBC-UK-2023
        \\ \hline\hline
    %
  %
  %
        parameter & method & value & Ref & source \\ \hline
        $\xi_0$ & indirect & \red{ $-1.738(177) \times 10^{-4}$ }
        & \cite{ RBC:2020kdj} & SWME \\
        $\xi_0$ & direct  & $-2.102(472) \times 10^{-4}$
        & \cite{ RBC:2020kdj} & SWME \\ \hline\hline
      \end{tabular}
    } 
    \caption{ Results for $\xi_0$ obtained using the direct and indirect
      methods in lattice QCD. }
    \label{tab:xi0-1}
  \end{subtable}
  \caption{Results for $\xi_0$. Here, we use the same notation as in
    Table \ref{tab:Vcb}. The abbreviation p4t1 means Table 1 on page
    4.}
  \label{tab:xi0-sum}
  \vspace*{-4mm}
\end{table}

%

%
%

%
%
\section{Input parameters: $\BK$,~ $\xi_\text{LD}$,~ and others}
The FLAG 2024 \cite{FlavourLatticeAveragingGroupFLAG:2024oxs} reports
results for $\BK$ in lattice QCD for $N_f=2$, $N_f=2+1$, and
$N_f=2+1+1$.
Here we use the results for $\BK$ with $N_f=2+1$, which is obtained by
taking an average over the five data points from BMW 11, Laiho 11,
RBC-UKQCD 14, SWME 14, and RBC-UKQCD 24 presented in Table
\ref{tab:input-BK-other} (\subref{tab:BK}).

\begin{table}[b!]
  \begin{subtable}{0.40\linewidth}
    \renewcommand{\arraystretch}{1.2}
    \resizebox{1.0\linewidth}{!}{
      \begin{tabular}{ l  l  l }
        \hline\hline
        Collaboration & Ref. & $\BK$  \\ \hline
	RBC/UKQCD 24  & \cite{Boyle:2024gge} & $0.7436(25)(78)$   \\
        SWME 15       & \cite{SWME:2015oos} & $0.735(5)(36)$     \\
        RBC/UKQCD 14  & \cite{RBC:2014ntl} & $0.7499(24)(150)$  \\
        Laiho 11      & \cite{Laiho:2011np} & $0.7628(38)(205)$  \\
        BMW 11        & \cite{BMW:2011zrh}  & $0.7727(81)(84)$  \\ \hline
        %
        FLAG-24       & \cite{FlavourLatticeAveragingGroupFLAG:2024oxs} p96e111    & \red{ $0.7533(91)$ } 
        \\ \hline\hline
      \end{tabular}
    } 
    \caption{$\BK$ 
	}
    \label{tab:BK}
  \end{subtable} 
  \hfill
  \begin{subtable}{0.57\linewidth}
    \renewcommand{\arraystretch}{1.2}
    \resizebox{1.0\linewidth}{!}{
      \begin{tabular}{ @{\qquad} c @{\qquad} l @{\qquad} l @{\qquad} }
        \hline\hline
        Input & Value & Ref. \\ \hline
        $G_{F}$
        & $1.1663788(6) \times 10^{-5} \GeV^{-2}$
        & PDG-24 \cite{ ParticleDataGroup:2024cfk} p137t1.1 \\ \hline
        $\theta$
        & $43.52(5)^{\circ}$
        & PDG-24 \cite{ ParticleDataGroup:2024cfk} p284e13.42\\ \hline
        $m_{K^{0}}$
        & $497.611(13)$ MeV
        & PDG-24 \cite{ ParticleDataGroup:2024cfk} p40 \\ \hline
        $\Delta M_{K}$
        & $3.484(6) \times 10^{-12}$ MeV
        & PDG-24 \cite{ ParticleDataGroup:2024cfk} p41 \\ \hline
        $F_K$
        & $155.7(3) \MeV$  
        & FLAG-24 \cite{FlavourLatticeAveragingGroupFLAG:2024oxs} p80e80\\ \hline
	$m_{c}(m_{c})$
	& \red{ $1.278(6) \GeV$ }
        & FLAG-24 \cite{FlavourLatticeAveragingGroupFLAG:2024oxs} p56e53 \\ \hline
	$m_{t}(m_{t})$
	& \red{ $162.77(27)(17) \GeV$ }
	& PDG-24 \cite{ ParticleDataGroup:2024cfk} p1379     \\ \hline
	$M_{W}$
	& \red{ $80.353(6) \GeV$ }
	& SM-2024 \cite{ParticleDataGroup:2024cfk} p815,s54p3 \\ \hline\hline 
      \end{tabular}
    } 
    \caption{Other parameters
		}
    \label{tab:other}
  \end{subtable} 
  \caption{ (\subref{tab:BK}) Results for $\BK$ and
    (\subref{tab:other}) other input parameters.}
  \label{tab:input-BK-other}

\end{table}

The dispersive long distance (LD) effect $\xi_\text{LD}$ is 
\begin{align}
  \xi_\text{LD} &=  \frac{m^\prime_\text{LD}}{\sqrt{2} \Delta M_K} \,,
  \qquad
  m^\prime_\text{LD}
  = -\Im \left[ \mathcal{P}\sum_{C}
    \frac{\mate{\wbar{K}^0}{H_\text{w}}{C} \mate{C}{H_\text{w}}{K^0}}
         {m_{K^0}-E_{C}}  \right]
  \label{eq:xi-LD}
\end{align}
There are two independent methods to estimate $\xi_\text{LD}$:
one is the BGI estimate \cite{Buras:2010pza}, and the other is the
RBC-UKQCD estimate \cite{Christ:2012se, Christ:2014qwa}, which
ex explained in Ref.~\cite{Bailey:2018feb}. 
In the BGI method, one estimates $\xi_\text{LD}$ using chiral
perturbation theory, using Eq.~\eqref{eq:xiLD:bgi}.
\begin{align}
  \xi_\text{LD} &= -0.4(3) \times \frac{\xi_0}{ \sqrt{2} }
  \label{eq:xiLD:bgi}
\end{align}
In the RBC-UKQCD method, one estimates $\xi_\text{LD}$ using
Eq.~\eqref{eq:xiLD:rbc}.
\begin{align}
  \xi_\text{LD} &= (0 \pm 1.6)\%
  \quad \text{of} \quad |\epsK|^\text{SM}.
  \label{eq:xiLD:rbc}
\end{align}
Here we use both methods to estimate the sie of $\xi_\text{LD}$.
In Table \ref{tab:input-BK-other} (\subref{tab:other}), we present other
input parameters needed to evaluate $\epsK$.
We present the charm quark mass $m_c(m_c)$ and the top quark mass
$m_t(m_t)$ in Table \ref{tab:input-BK-other} (\subref{tab:other}).
Since the lattice results of various groups with $N_f = 2+1+1$ shows
some inconsistency among them, we take the results for $m_c(m_c)$ with
$N_f = 2+1$ from FLAG 2024 \cite{
  FlavourLatticeAveragingGroupFLAG:2024oxs}.
To obtain $m_t(m_t)$, we take results for the pole mass $M_t$ from
PDG 2024 \cite{ ParticleDataGroup:2024cfk}.
Here we use the standard model predition (SM-2024) result for $M_W$
\cite{ ParticleDataGroup:2024cfk} to evaluate $\epsK$.
%

%
%



%
%
\section{Input parameters: Higher order QCD corrections}

We summarize higher order QCD corrections $\eta_i$ in 
Table \ref{tab:input-eta}.
There are two sets of $\eta_i$: one is $\eta_i$ of $c-t$ unitarity
(the traditional method \cite{Bailey:2015tba,Bailey:2018feb}, Table \ref{tab:input-eta}
(\subref{tab:eta_ctu})), and the other is $\eta_i$ of $u-t$ unitarity
(the BGS method \cite{Brod:2019rzc}, Table \ref{tab:input-eta}
(\subref{tab:eta_utu})).
\begin{table}[h!]
  \begin{subtable}{0.37\linewidth}
    \renewcommand{\arraystretch}{1.0}
    \resizebox{1.0\linewidth}{!}{
      \begin{tabular}[b]{ c @{\qquad} l @{\qquad} c }
        \hline\hline
        Input & Value & Ref.
        \\ \hline
        $\eta_{cc}$ & $1.72(27)$   & \cite{Bailey:2015tba}
        \\
        $\eta_{tt}$ & $0.5765(65)$ & \cite{Buras:2008nn}
        \\
        $\eta_{ct}$ & $0.496(47)$  &  \cite{Brod:2010mj}
        \\ \hline\hline
      \end{tabular}
    } 
    \caption{ $\eta_{i}$ of $c-t$ unitarity }
    \label{tab:eta_ctu}
  \end{subtable} 
  \hfill
  \begin{subtable}{0.61\linewidth}
    \renewcommand{\arraystretch}{1.56}
    \resizebox{1.0\linewidth}{!}{
      \begin{tabular}[b]{ c @{\qquad} l @{\qquad\qquad} @{\qquad} c }
        \hline\hline
        Input & Value & Ref.
        \\ \hline
        $\eta_{tt}^{\text{BGS}}$ & $0.55(1\pm4.2\%\pm0.1\%)$   & \cite{Brod:2019rzc}
        \\
        $\eta_{ut}^{\text{BGS}}$ & $0.402(1\pm1.3\%\pm0.2\%\pm0.2\%)$ & \cite{Brod:2019rzc}
        \\ \hline\hline
      \end{tabular}
    } 
    \caption{ $\eta_{i}^\text{BGS}$ of $u-t$ unitarity }
    \label{tab:eta_utu}
  \end{subtable} 
  \caption{ QCD corrections: (\subref{tab:eta_ctu}) the traditional
    method ($\eta_i$ of $c-t$ unitarity), and (\subref{tab:eta_utu})
    the BGS method ($\eta_i$ of $u-t$ unitarity). }
  \label{tab:input-eta}
\end{table}

The BGS method ($\eta_{ut}^\text{BGS}$) are supposed to have better
convergence with respect to the charm quark mass contribution
\cite{Brod:2019rzc}.
%

%
%

%
%
\section{Results for $|\epsK|^{\text{SM}}$}
Here we presents results for $|\epsK|^\text{SM}$ evaluated using
various combinations of input parameters.
We report results for $|\epsK|^\text{SM}_{c-t}$ calculated
using the traditional method with $\eta_{i}$ of $c-t$ unitarity in
Subsection ~\ref{sub:epsK:trad}, and
$|\epsK|^\text{SM}_{u-t}$ calculated using the BGS method
with $\eta_{i}$ of $u-t$ unitarity in Subsection ~\ref{sub:epsK:BGS}.
Here the superscript ${}^\text{SM}$ represents the theoretical
expectation value of $|\epsK|$ obtained directly from the SM, and the
subscript ${}_{c-t}$ (${}_{u-t}$) represents that obtained using the
traditional method with $\eta_i$ of $c-t$ unitarity
(the BGS method of $\eta_i$ of $u-t$ unitarity).

\subsection{$\eta_{i}$ of $c-t$ unitarity (the traditional method)}
\label{sub:epsK:trad}
\begin{figure}[t!]
  \begin{subfigure}{0.47\linewidth}
    \vspace*{-5mm}
    \includegraphics[width=1.0\linewidth]
       {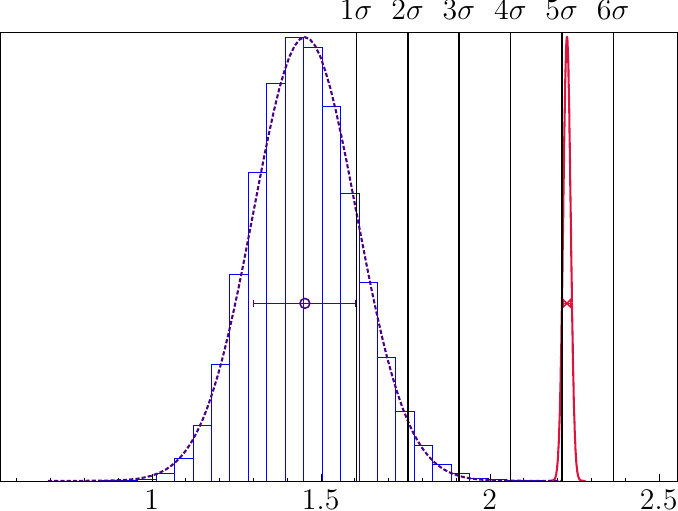}
    \caption{$|\epsK|^\text{SM}_{c-t}$}
    \label{fig:epsK-ex:ctu}
  \end{subfigure}
  \hfill
  \begin{subfigure}{0.47\linewidth}
    \vspace*{-5mm}
    \includegraphics[width=1.0\linewidth]
       {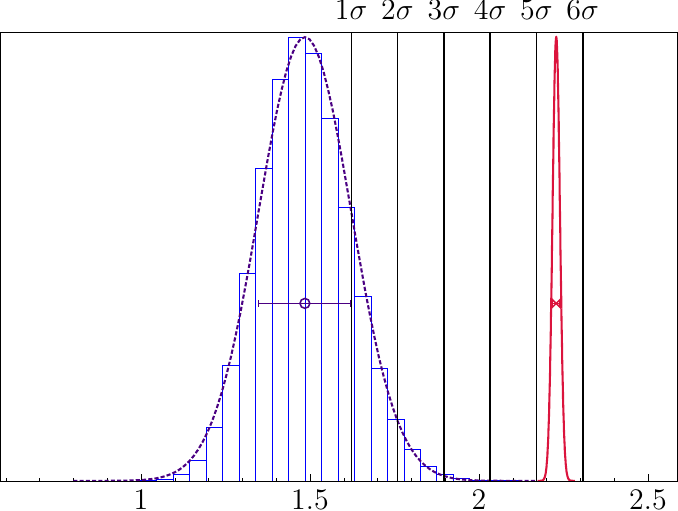}
    \caption{$|\epsK|^\text{SM}_{u-t}$}
    \label{fig:epsK-ex:utu}
  \end{subfigure}
  \caption{ $|\epsK|^\text{SM}$ with exclusive $\Vcb$ (FNAL/MILC-22)
    in units of $1.0 \times 10^{-3}$: (\subref{fig:epsK-ex:ctu})
    $|\epsK|^\text{SM}$ with $\eta_i$ of $c-t$ unitarity (the
    traditional method), and (\subref{fig:epsK-ex:utu})
    $|\epsK|^\text{SM}$ with $\eta_i$ of $u-t$ unitarity (the BGS
    method).  }
  \label{fig:epsK-ex:cmp}
    \vspace*{-6mm}
\end{figure}

In Fig.~\ref{fig:epsK-ex:cmp} (\subref{fig:epsK-ex:ctu}), we present
results of $|\epsK|^\text{SM}_{c-t}$ calculated directly from the
standard model (SM) with the lattice QCD inputs using the traditional
method for $\eta_i$ of $c-t$ unitarity.
Here the blue curve represents the theoretical results for
$|\epsK|^{\text{SM}}_{c-t}$ obtained using the FLAG-24 results for
$\BK$, the AOF results for Wolfenstein parameters, the FNAL/MILC-22
results for exclusive $\Vcb$, results for $\xi_0$ with the indirect
method, results for $\eta_i$ of $c-t$ unitarity (the traditional
method), and the RBC-UKQCD estimate for $\xi_\text{LD}$.
The red curve in Fig.~\ref{fig:epsK-ex:cmp} represents the
experimental results for $|\epsK|^\text{Exp}$.
Here the superscript ${}^\text{Exp}$ represents experimental results
for $|\epsK|$.
%
%
%

In Table \ref{tab:epsK}, we summarize our results for
$|\epsK|^\text{SM}_{c-t}$ and $\Delta\epsK = |\epsK|^\text{Exp} -
|\epsK|^\text{SM}$.
We present results for $|\epsK|^\text{SM}_{c-t}$ obtained using the
RBC-UKQCD estimate for $\xi_\text{LD}$ in Table \ref{tab:epsK}
(\subref{tab:epsK:rbc}), and those obtained using the BGI estimate for
$\xi_\text{LD}$ in Table \ref{tab:epsK} (\subref{tab:epsK:bgi}).
In Table \ref{tab:epsK}, we find that the theoretical expectation
values of $|\epsK|^\text{SM}_{c-t}$ with lattice QCD inputs (including
exclusive $\Vcb$) have $5.1\sigma \sim 4.1\sigma$ tension with the
experimental value of $|\epsK|^\text{Exp}$, while there is no tension
with inclusive $\Vcb$ obtained using the heavy quark expansion and QCD
sum rules.

\begin{table}[b!]
%
  \begin{subtable}{1.0\linewidth}
    \vspace*{-5mm}
    \center
    \renewcommand{\arraystretch}{1.1}
    \resizebox{0.99\linewidth}{!}{
      \begin{tabular}{l @{\qquad\qquad} l @{\qquad\qquad} l @{\qquad\qquad} l @{\qquad\qquad} l }
        \hline\hline
        $\Vcb$    & method   & source    & $|\epsK|^\text{SM}_{\text{$c-t$}}$ & $\Delta\epsK$
        \\ \hline
        exclusive & BGL      & FNAL/MILC-22 & $1.453 \pm 0.152$ & $5.10\sigma$
        \\
        exclusive & comb     & HFLAV-23     & $1.551 \pm 0.133$ & $5.10\sigma$
        \\        
        exclusive & comb     & FLAG-24      & $1.605 \pm 0.138$ & $4.50\sigma$
        \\
        exclusive & comb     & HPQCD-23     & $1.544 \pm 0.169$ & $4.06\sigma$
        \\ \hline
        inclusive & 1S       & 1S-23     & $2.017 \pm 0.155$ & $1.36\sigma$
        \\
        inclusive & kinetic  & Gambino-21   & $2.050 \pm 0.162$ & $1.10\sigma$
        \\ \hline\hline
      \end{tabular}
    } 
    \caption{$|\epsK|^{\text{SM}}_{\text{$c-t$}}$ with RBC-UKQCD estimate for $\xi_\text{LD}$}
    \label{tab:epsK:rbc}
  \end{subtable} 
  \begin{subtable}{1.0\linewidth}
    \vspace*{3mm}
    \center
    \renewcommand{\arraystretch}{1.1}
    \resizebox{0.99\linewidth}{!}{
      \begin{tabular}{ l @{\qquad\qquad} l @{\qquad\qquad} l @{\qquad\qquad} l @{\qquad\qquad} l }
        \hline\hline
        $\Vcb$    & method   & reference  & $|\epsK|^{\text{SM}}_{\text{$c-t$}}$ & $\Delta\epsK$
        \\ \hline
        exclusive & BGL  & FNAL/MILC-22 & $1.501 \pm 0.155$ & $4.70\sigma$
        \\
        exclusive & comb & HFLAV-23     & $1.599 \pm 0.135$ & $4.64\sigma$
        \\ \hline\hline
      \end{tabular}
    } 
    \caption{$|\epsK|^{\text{SM}}_{\text{$c-t$}}$ with BGI estimate for $\xi_\text{LD}$}
    \label{tab:epsK:bgi}
  \end{subtable} 
  \caption{ $|\epsK|$ in units of $1.0\times 10^{-3}$, and
    $\Delta\epsK = |\epsK|^\text{Exp} - |\epsK|^{\text{SM}}_{\text{$c-t$}}$ in units of
    $\sigma$.}
  \label{tab:epsK}
\end{table}

\begin{figure}[t!]
  \begin{subfigure}{0.501\linewidth}
    \vspace*{-6mm}
    \includegraphics[width=\linewidth]
                    {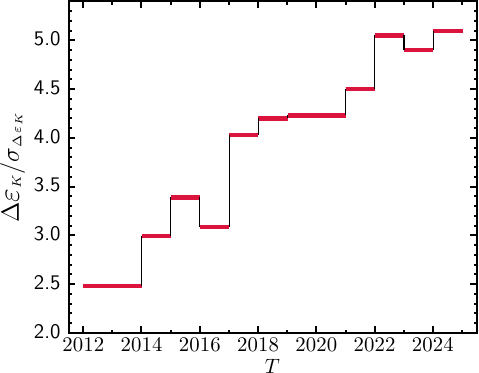}
    \caption{Time evolution of $\Delta \epsK/\sigma_{\Delta \epsK}$}
    \label{fig:depsK:rbc:his}
  \end{subfigure}
  \hfill
  \begin{subfigure}{0.459\linewidth}
    \vspace*{-6mm}
    \includegraphics[width=\linewidth]
                    {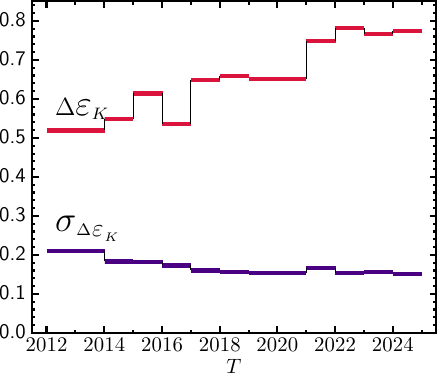}
    \caption{Time evolution of the average and error of $\Delta\epsK$}
    \label{fig:depsK+sigma:rbc:his}
  \end{subfigure}
  \vspace*{-2mm}
  \caption{ Chronicle of (\subref{fig:depsK:rbc:his})
    $\Delta\epsK/\sigma_{\Delta \epsK}$, and
    (\subref{fig:depsK+sigma:rbc:his}) $\Delta\epsK$ and
    $\sigma_{\Delta\epsK}$. }
  \label{fig:depsK:sum:rbc:his}
  \vspace*{-4mm}
\end{figure}

In Fig.~\ref{fig:depsK:sum:rbc:his} (\subref{fig:depsK:rbc:his}), we
present the time evolution of $\Delta\epsK/\sigma$ during the period
of 2012--2024.
In 2012, $\Delta\epsK$ was $2.5\sigma$, but now it is $5.1\sigma$ with
exclusive $\Vcb$ (FNAL/MILC-22).
We use the FNAL/MILC-22 results for exclusive $\Vcb$ as a
representative sample, since it contains the most comprehensive
analysis of the $\BtoDst$ decays at both zero recoil and non-zero
recoil, while it incorporates experimental results of both BELLE and
BABAR, and independent of data merging with unwanted correlation.
In Fig.~\ref{fig:depsK:sum:rbc:his} (\subref{fig:depsK+sigma:rbc:his})
we present the time evolution of the average $\Delta\epsK$ and the error
$\sigma_{\Delta\epsK}$ during the same period.
%
%
%

In Table \ref{tab:err-budget} (\subref{tab:err-budget:ctu:rbc}),
we present the error budget for $|\epsK|^\text{SM}_{c-t}$.
Here we find that $\Vcb$ gives the largest error ($\approx 52\%$),
while $\eta_{ct}$, $\bar\eta$, and $\eta_{cc}$ are subdominant in the
error budget.
Hence, it is essential to reduce the errors in $\Vcb$ as much as
possible.
Part of the errors in exclusive $\Vcb$ come from experiments in BELLE,
BELLE2, BABAR, and LHCb, which are beyond our control, but will
decrease thanks to on-going accumulation of higher statistics in
BELLE2 and LHCb.
Part of the errors in exclusive $\Vcb$ come from the theory used to
evaluate the semi-leptonic form factors for $\BtoDstp$ decays, using
tools in lattice QCD.
%
%
%
\begin{table}[b!]
  \begin{subtable}{0.48\linewidth}
    \center
    \vspace*{-5mm}
    \renewcommand{\arraystretch}{1.05}
    \resizebox{1.0\linewidth}{!}{
      \begin{tabular}{ @{\quad} c @{\qquad\qquad} l @{\qquad\qquad} l
          @{\quad} }
	\hline\hline source & error (\%) & memo \\ \hline $\Vcb$ &
        51.9 & exclusive \\ $\eta_{ct}$ & 21.9 & $c-t$ Box
        \\ $\bar{\eta}$ & \phantom{0}9.4 & AOF \\ $\eta_{cc}$ &
        \phantom{0}9.3 & $c-c$ Box \\ $\xi_\text{LD}$ & \phantom{0}2.2
        & RBC/UKQCD \\ $\bar{\rho}$ & \phantom{0}2.0 & AOF
        \\ $\hat{B}_K$ & \phantom{0}1.6 & FLAG-24 \\
        $\vdots$        & \phantom{00}$\vdots$  & $\vdots$ \\
        \hline\hline
    \end{tabular}}
    \caption{ Error budget for $|\epsK|^\text{SM}_{c-t}$}
    \label{tab:err-budget:ctu:rbc}
  \end{subtable}
  \hfill
  \begin{subtable}{0.48\linewidth}
    \center
    \vspace*{-5mm}
    \renewcommand{\arraystretch}{1.05}
    \resizebox{1.0\linewidth}{!}{
      \begin{tabular}{ @{\quad} c @{\qquad\qquad} l @{\qquad\qquad} l @{\quad}}
	\hline\hline
        source          & error (\%)            & memo \\
        \hline
        $\Vcb$                   &           63.1        & exclusive \\
        $\bar{\eta}$             &           12.0        & AOF \\
        $\eta_{tt}^\text{BGS}$   &           10.7        & BGS \\
        $\delta\epsK^\text{BGS}$ & \phantom{0}5.4        & CV mismatch \\
        $\xi_\text{LD}$          & \phantom{0}2.9        & RBC/UKQCD \\
        $\bar{\rho}$             & \phantom{0}2.2        & AOF \\
        $\hat{B}_K$              & \phantom{0}2.0        & FLAG-24 \\
        $\vdots$                 & \phantom{00}$\vdots$  & $\vdots$ \\
	\hline\hline
    \end{tabular}}
    \caption{ Error budget for $|\epsK|^\text{SM}_{u-t}$}
   \label{tab:err-budget:utu:rbc}
  \end{subtable}
  \caption{ Error budget table for $|\epsK|^\text{SM}$ with
    (\subref{tab:err-budget:ctu:rbc}) the traditional method ($\eta_i$
    of $c-t$ unitarity), and (\subref{tab:err-budget:utu:rbc}) the BGS
    method ($\eta_i$ of $u-t$ unitarity).  }
  \label{tab:err-budget}
  \vspace*{-7mm}
\end{table}

\subsection{$\eta_{i}$ of $u-t$ unitarity (the BGS method)}
\label{sub:epsK:BGS}
In Fig.~\ref{fig:epsK-ex:cmp} (\subref{fig:epsK-ex:utu}), we present
results for $|\epsK|^\text{SM}_{u-t}$ obtained directly from the SM
using the BGS method for $\eta_i$ of $u-t$ unitarity, the FLAG-24
results for $\BK$, the AOF results for Wolfenstein parameters, the
FNAL/MILC-22 results for exclusive $\Vcb$, results for $\xi_0$ with
the indirect method, the RBC-UKQCD estimate for $\xi_\text{LD}$, and
so on.
In Table \ref{tab:epsK:utu:rbc:indirect}, we present results for
$|\epsK|^\text{SM}_{u-t}$ and its $\Delta \epsK$ obtained using
the BGS method.
Here we find a mismatch $\delta \epsK^\text{BGS}$ in the central
values (CV) for $|\epsK|^\text{SM}$ between the traditional method
and the BGS method for $\eta_i$: $\delta \epsK^\text{BGS} \equiv
  |\epsK|^\text{SM}_{u-t} - |\epsK|^\text{SM}_{c-t} $.
This mismatch comes from a number of small and tiny approximations
introduced in the BGS method \cite{Brod:2019rzc}.
Here we count this CV mismatch $\delta \epsK^\text{BGS}$ as an
additional error, and add it to the total error in quadrature.
Hence, the total error is $ \sigma_t^\text{BGS} = \sqrt{
  [\sigma_1^\text{BGS}]^2 + [\delta \epsK^\text{BGS}]^2} $, where
$\sigma_t^\text{BGS}$ represents the total error, and
$\sigma_1^\text{BGS}$ represents the errors coming from the input
parameters.
In Table \ref{tab:epsK:utu:rbc:indirect}, we present
$\sigma_1^\text{BGS}$, $\delta\epsK^\text{BGS}$, and
$\sigma_t^\text{BGS}$ to demonstrate some sense on numerical size of
them.
From Table \ref{tab:epsK:utu:rbc:indirect}, we find that the
theoretical results for $|\epsK|^\text{SM}_{u-t}$ obtained using
lattice QCD inputs including exclusive $\Vcb$ have $5.7\sigma \sim
4.2\sigma$ tension with the experimental results for
$|\epsK|^\text{Exp}$, while the tension disappears for those obtained
using inclusive $\Vcb$ from heavy quark expansion and QCD sum rules.
%
%
%
\begin{table}[t!]
  \center
  \renewcommand{\arraystretch}{1.1}
  \vspace*{-5mm}
  \resizebox{1.0\linewidth}{!}{
    \begin{tabular}{ l @{\qquad} l @{\qquad}
        l @{\qquad} l @{\qquad}
        l @{\qquad} l @{\qquad} l @{\qquad} r }
      \hline\hline
      $\Vcb$    & method   & source       & $|\epsK|^\text{SM}_{\text{$u-t$}}$
      & $\sigma_1^\text{BGS}$ & $\delta\epsK^\text{BGS}$
      & $\sigma_t^\text{BGS}$ & $\Delta\epsK/\sigma$
      \\ \hline
      excl      & BGL      & FNAL-MILC-22 & 1.484  & 0.133     & 0.032  & 0.137  &5.43      \\
      excl      & comb      & HFLAV-23 & 1.582  & 0.110     & 0.031  & 0.114  &5.65      \\
      excl      & comb      & FLAG-24 & 1.635   & 0.116     & 0.030  & 0.120  &4.93      \\
      excl      & comb      & HPQCD-23 & 1.575  & 0.151     & 0.031  & 0.154  &4.24      \\
      \hline
      incl      & 1S      & 1S-23 & 2.043 & 0.131     & 0.026  & 0.134  &1.37      \\
      incl      & kinetic      & Gambino-21 & 2.075   & 0.140     & 0.025  & 0.142  &1.07    
      \\ \hline
      $|\epsK|^\text{Exp}$
      & exp      & PDG-24                 & 2.228 &       &       & 0.011
      & $0.00$ \\\hline\hline
    \end{tabular}
  }
  \caption{ $|\epsK|^\text{SM}_{u-t}$ in units of $1.0 \times 10^{-3}$
    obtained using the FLAG-24 results for $\BK$, AOF for the
    Wolfenstein parameters, the \texttt{indirect} method for $\xi_0$,
    the \texttt{RBC-UKQCD} estimate for $\xi_\mathrm{LD}$, and the
    \texttt{BGS} method for $\eta_i$ ($=\eta_i^\text{BGS}$) of $u-t$
    unitarity. }
  \label{tab:epsK:utu:rbc:indirect}  
\end{table}

In Table \ref{tab:err-budget} (\subref{tab:err-budget:utu:rbc}),
we present the error budget for $|\epsK|^\text{SM}_{u-t}$.
Here we find that the error from exclusive $\Vcb$ is dominant
($\approx 63\%$), while those errors from $\bar\eta$,
$\eta^\text{BGS}_{tt}$, and $\delta\epsK^\text{BGS}$ are subdominant
in the error budget.
Hence, it is essential to reduce the errors of exclusive $\Vcb$
as much as possible.

Due to lack of space, a large portion of interesting results for
$|\epsK|^\text{SM}$ and $\Delta\epsK$ could not be presented in Tables
\ref{tab:epsK} and \ref{tab:epsK:utu:rbc:indirect}: for example,
results for $|\epsK|^\text{SM}_{c-t}$ obtained using exclusive $\Vcb$
(FLAG-24) with the BGI estimate for $\xi_\text{LD}$, results for
$|\epsK|^\text{SM}_{c-t}$ obtained using $\xi_0$ determined by the
direct method, and so on.
We plan to report them collectively in Ref.~\cite{ wlee:2024epsK}.

In order to reduce the errors in exclusive $\Vcb$ on the theoretical
side, there is an on-going project to determine $\Vcb$ using
the Oktay-Kronfeld (OK) action for the heavy quarks to calculate the
form factors for $\BtoDstp$ decays \cite{ Bhattacharya:2021peq,
  Park:2020vso, Bhattacharya:2020xyb, LANLSWME:2018dde,
  Bailey:2017xjk, Bailey:2017zgt, Bailey:2020uon,
  Bhattacharya:2023xgf, LANL-SWME:2023gow}.

%
%

\acknowledgments
We thank J.~Bailey, Y.C.~Jang, S.~Sharpe, and R.~Gupta for helpful
discussion.
We thank G.~Martinelli for providing us with the most updated results
of the UTfit Collaboration in time.
The research of W.~Lee is supported by the Mid-Career Research
Program (Grant No.~NRF-2019R1A2C2085685) of the NRF grant funded by
the Korean government (MSIT).
W.~Lee would like to acknowledge the support from the KISTI
supercomputing center through the strategic support program for the
supercomputing application research (No.~KSC-2018-CHA-0043,
KSC-2020-CHA-0001, KSC-2023-CHA-0010, KSC-2024-CHA-0002).
Computations were carried out in part on the DAVID cluster at Seoul
National University.
%

\bibliography{refs}


\end{document}